\documentclass[letterpaper, 10 pt, conference]{ieeeconf}  

\IEEEoverridecommandlockouts                              

\usepackage{cite}
\usepackage{amsmath,amssymb,amsfonts}
\usepackage{booktabs}
\usepackage{algorithm}
\usepackage{graphicx}
\usepackage{textcomp}
\usepackage{xcolor}
\usepackage{float}
\usepackage{balance}
\usepackage{algpseudocode}
\usepackage{amsmath, amsfonts}
\usepackage{makecell}

\title{\LARGE \bf
AttentionSwarm: Reinforcement Learning with Attention Control Barrier Function for Crazyflie Drones in Dynamic Environments
}

\author{Grik Tadevosyan, Valerii Serpiva, Aleksey Fedoseev, Roohan Ahmed Khan, Demetros Aschu, Faryal Batool, \\
Nickolay Efanov, Artem Mikhaylov, and Dzmitry Tsetserukou
\thanks{Grik Tadevosyan is with the Mohamed bin Zayed University of Artificial Intelligence (MBZUAI), Abu Dhabi, United Arab Emirates. {\tt \{Grik.Tadevosyan\}@mbzuai.ac.ae}}%
\thanks{Valerii Serpiva, Aleksey Fedoseev, Roohan Ahmed Khan, Demetros Aschu, Faryal Batool, and Dzmitry Tsetserukou are with the Intelligent Space Robotics Laboratory,
Skolkovo Institute of Science and Technology, Moscow, Bolshoy Boulevard 30, bld. 1, 121205, Moscow, Russia.
{\tt \{Valerii.Serpiva, Aleksey.Fedoseev, Roohan.Khan, Demetros.Tareke, Faryal.Batool, D.Tsetserukou\}}@skoltech.ru}%
\thanks{Nickolay Efanov and Artem Mikhaylov are with the AI Center, Skolkovo Institute of Science and Technology, Moscow, Russia. {\tt \{N.Efanov, A.Mikhaylov\}}@skoltech.ru}%
\thanks{Nickolay Efanov is also with the Moscow Institute of Physics and Technology (National State Univ.), Department of Informatics and Computational Mathematics, Dolgoprudny, Russia. {\tt efanov.nn@mipt.ru}}%
}

\begin{document}

\maketitle
\thispagestyle{empty}
\pagestyle{empty}

\begin{abstract}
We introduce \textit{AttentionSwarm}, a novel benchmark designed to evaluate safe and efficient swarm control in a dynamic drone racing scenario. Central to our approach is the Attention Model-Based Control Barrier Function (CBF) framework, which integrates attention mechanisms with safety-critical control theory to enable real-time collision avoidance and trajectory optimization. This framework dynamically prioritizes critical obstacles and agents in the swarm’s vicinity using attention weights, while CBFs formally guarantee safety by enforcing collision-free constraints. The AttentionSwarm algorithm was developed and evaluated using a swarm of Crazyflie 2.1 micro quadrotors, which were tested indoors with the Vicon motion capture system to ensure precise localization and control. Experimental results show that our system achieves a 95-100\% collision-free navigation rate in a dynamic multi-agent drone racing environment, underscoring its effectiveness and robustness in real-world scenarios. This work offers a promising foundation for safe, high-speed multi-robot applications in logistics, inspection, and racing.

\end{abstract}

\section{Introduction}

In recent years, Deep Reinforcement Learning (DRL) \cite{Tang_Abbatematteo_Hu_Chandra_Martín-Martín_Stone_2025} has emerged as a critical methodology in robotics, driving advances in systems that require adaptability \cite{peter2024, huang2024, petertornado}. However, ensuring system safety remains paramount. A promising approach is the implementation of safe learning techniques \cite{wan2020}, \cite{tao2022} and Safe Reinforcement Learning (Safe RL) \cite{feng2025}, \cite{xu2025}. Nevertheless, Safe RL methods have limitations: in dynamic environments with moving obstacles, these methods typically do not retain or utilize information about the obstacles' states, which is crucial for accurately predicting subsequent states. Although the approaches, e.g., Model Predictive Control (MPC) combined with safe learning, offer certain safety guarantees, they often fall short in dynamic and complex environments. MPC-based methods can struggle with slow adaptation times, which is problematic when the environment is replete with unknown and moving obstacles, such as pedestrians, where accurate position data are not available. In scenarios with a high number of states or complex environments, the performance of MPC can substantially degrade, leading to safety risks \cite{safari2024},\cite{zhao2023}.
A key challenge for MPC is scalability with the number of agents. For MPC, each agent should know the dynamics of other agents and the environment; however, in real problems, agents do not know the dynamics of other agents and the environment \cite{lapandic2021}. 

Furthermore, as the complexity of multi-agent systems grows, many algorithms exhibit scalability issues \cite{andrew2022}.
One important theoretical work on the Safe RL is presented in \cite{gu2023}. However, the authors have not yet validated their approach through real-world experiments.

We propose a novel algorithm that enhances Multi-Agent Proximal Policy Optimization (MAPPO) by integrating a Multi-Layer Perceptron (MLP) Neural Network (NN) architecture and enhances it with a Control Barrier Net Function that utilizes attention models to predict the optimal parameters for the Control Barrier Function (CBF). This approach is specifically designed for centralized multi-agent systems operating in environments with moving obstacles, enabling more precise state estimation and robust safety guarantees.

\begin{figure}
    \centering
    \includegraphics[width=0.95\linewidth]{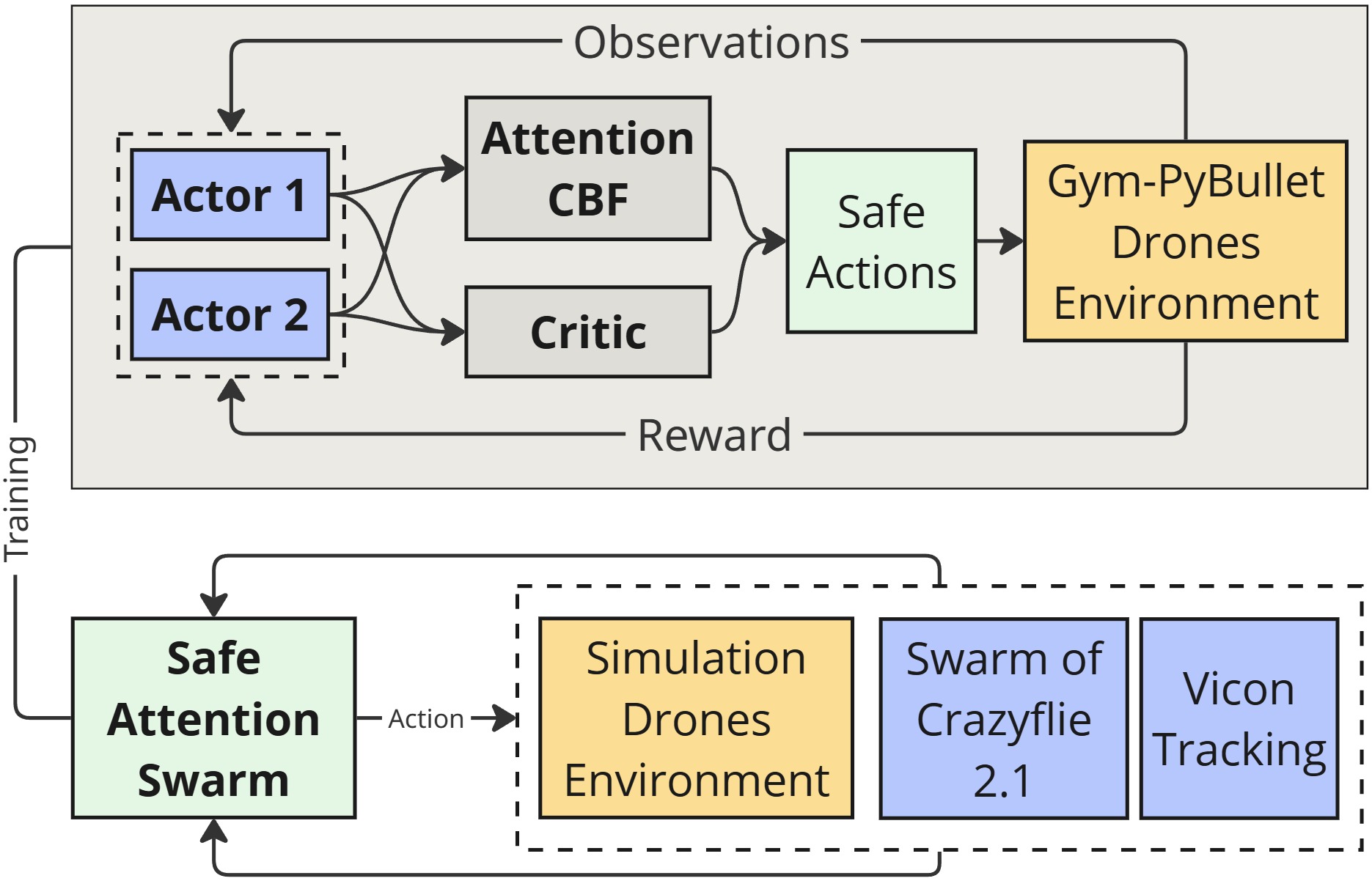}
    \caption{System overview of AttentionSwarm, a reinforcement learning-based system enabling drone swarms to safely navigate in dynamic environments with moving obstacles and changing conditions. The system is evaluated across a dynamic drone racing scenario.}
    \label{fig:sys_over}
\end{figure}

\section{Related Work}
In recent years, DRL has advanced robotics by enabling systems to adapt to complex, dynamic environments \cite{bartolomei2022, wang2024, zhao2024, roohan2025}. 
Recent advancements in Safe Reinforcement Learning (Safe RL) have produced several impressive projects designed for dense environments \cite{yifru2024, liu2024}. These approaches have demonstrated robust safety guarantees during exploration and deployment in complex settings. However, many of these methods exhibit limitations when extended to multi-agent systems. In particular, agents in these frameworks face memory challenges, when they do not adequately store or leverage previous obstacle states, which is critical for dynamic environments.

Moreover, while there are remarkable papers on Safe RL, some of these contributions focus on single-agent scenarios \cite{Amendola2024, srinivasan2020, yang2022, Xie2023, zhou2023, tadevosyan2025}. Such studies, although effective in ensuring safety for individual agents, do not address the unique challenges that arise when multiple agents operate concurrently, especially in terms of memory and state retention.
There are also works in drone racing using DRL \cite{Loquercio_2020}, \cite{geles2024}, yet these approaches have not been extended to multi-agent systems.

Remarkable projects have been developed for single-agent dynamic systems that integrate computer vision to capture the velocity and moving direction of obstacles \cite{xu2025}, effectively transforming simulation models into real-world applications. Although these methods harness DRL to achieve impressive performance, they remain confined to single-agent contexts.
Another method explored in the literature leverages Graph Neural Networks \cite{zhang2025} in conjunction with CBF to determine optimal CBF parameters for swarms of drones in dynamic environments. While this approach has shown potential in coordinating multiple agents, it suffers from a significant drawback: a higher adaptation time.
One remarkable work is \cite{zhang2024}, where the authors employed Safe RL with hierarchical control for a multi-agent system. However, this approach exhibits two major limitations: computational complexity, scalability, and limited real-world applications. 
Other approaches in the literature have explored the use of classical safe learning and MPC-based models \cite{fisac2019} \cite{mao2023}. While these methods have shown promise under controlled conditions, they exhibit substantial limitations under complex and dynamic environments with disturbances. Moreover, when applied to multi-agent systems, these models do not perform satisfactorily due to their inability to scale effectively. Recent work has also combined MPC with attention-based models \cite{jacquet2024}, where the parameters of MPC are determined using NNs; however, such approaches have been demonstrated only for single-drone scenarios.
One other work is \cite{xiao2023}, where the authors employed NNs in conjunction with CBF to enforce safety constraints. However, a notable limitation of this approach is that it requires prior knowledge of the environment's dynamic properties. This dependency restricts its applicability in scenarios where the dynamics are uncertain or rapidly changing. In contrast, a Nonlinear Model Predictive Control \cite{11097824} with obstacle avoidance enables rapid and safe drone navigation based on the multimodal system’s output.

In this work, we propose a novel algorithm that combines multi-agent DRL with control barrier net functions enhanced by attention models. By incorporating an attention mechanism, our approach effectively compensates for memory limitations by recalling historical obstacle information, enabling better prediction of future states. Our method has demonstrated promising results across a swarm navigation and a drone racing environment, highlighting its potential for robust safety in multi-agent, dynamic settings.

\section{AttentionSwarm Methodology}

\subsection{System Overview}

The AttentionSwarm system depicted in Fig. \ref{fig:sys_over} comprises a swarm of Bitcraze Crazyflie drones operating in an environment with a moving obstacle and a static gate. A Vicon Vantage V5 motion capture system is utilized for microdrone localization, transmitting updated position and orientation data via the Robot Operating System to the control station. The control station processes this information to estimate linear and angular velocities, which are then forwarded to the model for the generation of safe policies. The safe policy produces control outputs at each time step, which are subsequently relayed to the Crazyflie onboard controller for motor actuation. This closed-loop process generates safe trajectories for each drone, guiding them to pass the gate successfully.
In dynamic environments with obstacles, ensuring the safety of our system is paramount. 

In our work, we address the challenge of safe navigation in the presence of moving obstacles by leveraging a centralized multi-agent Proximal Policy Optimization (PPO) algorithm augmented with a CBF Attention Neural Network. This integration allows our model to actively enforce safety constraints during training and adapt to changing obstacle positions.

\begin{figure}
    \centering
    \includegraphics[width=0.8\linewidth]{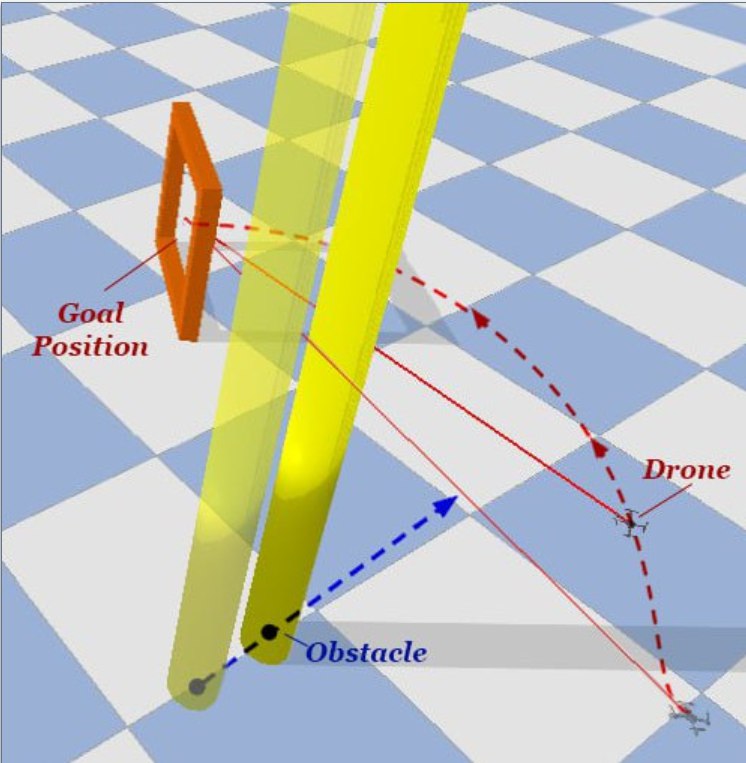}
    \caption{Gym-PyBullet simulation setup with a moving obstacle.}
    \label{fig:moving_obs}
\end{figure}

Our centralized multi-agent PPO framework, integrated with the CBF Attention Neural Network, enables the drones to learn safe policies that effectively handle the presence of dynamic obstacles, thereby enhancing both safety and performance across various scenarios.

\subsection{Simulation Environment} 

The simulation environment was developed using the \texttt{gym-pybullet-drones} framework \cite{panerati2021}. As illustrated in Fig.~\ref{fig:moving_obs}, our custom environment comprises a swarm of Crazyflie drones, static and dynamic obstacles, and navigation gates, modeled as cylinders and rectangles. All training was conducted on a workstation equipped with an Intel Core i9 CPU, 64 GB of RAM, and an NVIDIA GeForce RTX 4090 GPU with 24 GB of VRAM. For real-world validation, a Vicon motion capture system provided ground-truth pose data with sub-millimeter accuracy, essential for high-speed drone control.

\subsection{Problem Formulation and Preliminaries}\
The multi-agent drone racing problem is addressed using state-of-the-art DRL techniques combined with Attention-based CBF. The Algorithm~\ref{alg:Centralized_MAPPO_CBF} describes the process in detail. In a 3D environment with \(N\) quadrotor drones, the problem is modeled as a Markov Decision Process (MDP) denoted as:

\begin{equation}
\text{MDP}(O_t, A_t, R_t),
\end{equation}

where \(O_t\) is the observation of the environment and the drone’s state, including kinematic information, along with information about static and moving obstacles, \(A_t\) is actions, \(R_t\) is the reward signal, which guides our objective.
The aim is to develop a safe policy using Attention Models for CBF from each drone to control actions that drive it to the gate while preventing collisions among drones and obstacles. 
The observation vector of the \(i\)th drone in the swarm is denoted by:

\begin{equation}
O_t^i = \Big[p_t^i, \; q_t^i, \; v_t^i, \; \omega_t^i, \; \text{obs}_t\Big],
\end{equation}

where
\(p_t^i\) represents the drone's state relative to the gate center point,
\(q_t^i\) denotes the drone's orientation, \(v_t^i\) signifies the velocity of the drone,  \(\omega_t^i\) indicates the angular velocity and \(\text{obs}_t\) represents the positions of obstacles.

\makeatletter
\algrenewcommand\algorithmicindent{0.9em}
\makeatother

\begin{algorithm}[!t]
\caption{Centralized MAPPO with CBF Constraint using Attention Models}
\label{alg:Centralized_MAPPO_CBF}
\begingroup
\small
\setlength{\abovedisplayskip}{2pt}
\setlength{\belowdisplayskip}{2pt}
\setlength{\abovedisplayshortskip}{1pt}
\setlength{\belowdisplayshortskip}{1pt}

\begin{algorithmic}[1]
\State \textbf{Initialize:} Centralized actor network $\pi_{\theta}$, centralized value network $V_{\phi}$, $CBF_{\psi}$
\State \textbf{Initialize:} Target networks $\pi_{\theta'}$, $V_{\phi'}$, and $CBF_{\psi'}$ with parameters set as $ \theta' \leftarrow \theta,\ \phi' \leftarrow \phi,\ \psi' \leftarrow \psi$
\For{each episode}
   \State Initialize the joint state $s_0$ for all agents
    \For{each timestep $t$}
        \State For each agent, sample action $a_t \sim \pi_{\theta}(s_t)$
        \State Execute $a_t$; observe $r_t$, $s_{t+1}$, and $obs_t$
        \State \textbf{Attention Mechanism for CBF:}
        \For{each obstacle $i$}
            \State Extract obstacle feature $x_i$
            \State Compute key $K_i = W_K x_i$
            \State Compute value $V^{att}_i = W_V^{att} x_i$
        \EndFor
        \State Compute query $Q = W_Q h$
        \State Compute attention weights for each obstacle $i$:
        \Statex \(\displaystyle
           \alpha_i = \frac{\exp\!\left((Q \cdot K_i)/\sqrt{d}\right)}
           {\sum_j \exp\!\left((Q \cdot K_j)/\sqrt{d}\right)}
        \)
        \Statex where $d$ is the dimension of the key vectors
        \State Compute context vector:
        \Statex \(\displaystyle c = \sum_i \alpha_i V^{att}_i \)
        \State Compute optimal CBF parameters:
        \Statex \(\displaystyle \theta_{CBF} = f_{\text{CBF}}(h, c) \)
        \Statex where \(f_{\text{CBF}}\) is a feed-forward network
        \State \textbf{Store transition} $(s_t, a_t, r_t, s_{t+1}, \theta_{CBF})$ in replay buffer $D$
    \EndFor

    \State \textbf{Sample a mini-batch} of transitions from $D$
    \State \textbf{Compute PPO Loss:}
    \Statex \(\displaystyle
       r_t(\theta) = \frac{\pi_{\theta}(a_t|s_t)}{\pi_{\theta_{\text{old}}}(a_t|s_t)}
    \) \ Statex and the clipped surrogate loss:
    \Statex \(\displaystyle
       L_{\text{PPO}}(\theta) =
       \mathbb{E}\!\left[\min\!\Big(r_t(\theta)\hat{A}_t,\,
       \operatorname{clip}(r_t(\theta),1-\epsilon,1+\epsilon)\hat{A}_t\Big)\right]
    \)
    \State \textbf{Compute Value Loss:}
    \Statex \(\displaystyle
       L_V(\phi) = \mathbb{E}\!\left[(V_\phi(s_t) - V_t^{\text{target}})^2\right]
    \)
    \State \textbf{Compute CBF Loss (Constraint):}
    \Statex \(\displaystyle
       L_{\text{CBF}}(\psi) = \mathbb{E}\!\left[\|CBF_{\psi}(s_t,a_t) -
       \theta_{CBF}^{\text{target}}\|^2\right]
    \)
    \Statex where $\theta_{CBF}^{\text{target}}$ is the desired optimal CBF parameter
    \State \textbf{Total Loss:} Update $\theta$, $\phi$, and $\psi$ by minimizing:
    \Statex \(\displaystyle
       L = L_{\text{PPO}}(\theta) + c_1 L_V(\phi) + c_2 L_{\text{CBF}}(\psi),
       \quad c_1,c_2 \ge 0
    \)
    \State \textbf{Update Target Networks:}
    \Statex \(\displaystyle
       \theta' \leftarrow \tau \theta + (1-\tau)\theta',\quad
       \phi' \leftarrow \tau \phi + (1-\tau)\phi',\quad
       \psi' \leftarrow \tau \psi + (1-\tau)\psi'
    \)
    \Statex where $\tau$ is the soft update coefficient
\EndFor
\end{algorithmic}
\endgroup
\end{algorithm}

\begin{equation}
    A_t^i = \Big[v_{x,t}^i, \; v_{y,t}^i, \; v_{z,t}^i\Big],
\end{equation}

where \(v_{x,t}^i\), \(v_{y,t}^i\), and \(v_{z,t}^i\) denote the velocities along the \(x\), \(y\), and \(z\) axes, respectively. These velocity commands are subsequently used to compute the drone's next position as it moves toward the center of the gate.
The reward signal for the \(i\)th drone, denoted by \(R_t^i\), is defined as:

\begin{equation}
R = \sum_{i} r_i,
\end{equation}

where

\begin{equation}
r_i =\begin{cases}
\frac{1}{d_i + \epsilon} & \text{if no collision,} \\
-100 & \text{if collision,}
\end{cases}
\end{equation}

and

\begin{equation}
d_i = \| p_i - p_{t} \|_2,
\end{equation}

where \(p_i\) representing the coordinates of the \(i\)th drone, \(p_{t}\) is the center of the gate point for the \(i\)th drone, and \(\epsilon = 0.001\) is the small constant added for stabilization to avoid division by zero.

\subsection{Centralized MAPPO with Attention CBF}

The MAPPO-Lagrangian algorithm (MAPPO with Lagrangian Multipliers) \cite{ding2023} is a model-free safety-aware Multi-Agent RL algorithm. It extends the PPO framework to multi-agent settings while incorporating soft safety constraints expressed as cost functions. In the standard formulation, the algorithm is designed to solve a constrained Markov game where each agent \(i\) is associated with a set of cost functions \(C_{ij}\) and corresponding safety constraints \(c_{ij}\). The goal is to maximize the expected cumulative reward \(J(\pi)\) while ensuring that the expected cumulative costs \(J_{ij}(\pi)\) do not exceed the safety thresholds \(c_{ij}\).

In our modified approach, we replace the traditional Lagrangian multipliers with an Attention-based CBF Neural Network, denoted as \(CBF_{\psi}\). Instead of using fixed multipliers, the network dynamically computes attention weights that serve as safety penalties. These weights are applied to the cost advantage functions, effectively enforcing the safety constraints as a soft constraint. This modification is particularly beneficial in dynamic environments, where moving obstacles and varying conditions require adaptive safety measures.

The optimization problem for each agent \(i\) is reformulated as a min-max problem. The modified objective function for the agent \(i\) becomes:

\begin{equation}
    \begin{split}
        \max_{\theta_i} \min_{\psi} \; \mathbb{E}_{s \sim \rho_{\pi_{\theta_k}}, \, a \sim \pi_{\theta_i}} & \left[ A_i^{\pi_{\theta_k}}(s,a) \right. \\
        \left. - \sum_{j=1}^{m_i} \theta_i \left( A_{ij,\pi_{\theta_k}}(s,a_i) + d_{ij} \right) \right],
    \end{split}
\end{equation}
where 
\vspace{-1em}
\begin{equation}
d_{ij} = J_{ij}(\pi_{\theta_k}) - c_{ij}, \; \theta_i = CBF_{\psi}(s, a_i)
\end{equation}

\(A_i^{\pi_{\theta_k}}(s,a)\) is the advantage function for the agent \(i\), and \(A_{ij,\pi_{\theta_k}}(s,a_i)\) is the cost advantage function for the \(j\)th constraint. The term \(CBF_{\psi}(s, a_i)\) outputs an attention-based weight that dynamically penalizes deviations from the safety constraints.

Both the policy parameters \(\theta_i\) and the Attention CBF network parameters \(\psi\) are updated iteratively. The policy is updated via gradient ascent on the modified PPO objective, while the Attention CBF network is updated using gradient descent on the penalty term. These updates are performed within a trust region defined by a divergence constraint Kullback-Leibler, ensuring stable and reliable policy updates.

This theoretical formulation guaranties that our method maximizes cumulative rewards while ensuring adherence to predefined safety constraints. The dynamic attention mechanism in the CBF network allows our algorithm to effectively manage the trade-off between performance and safety in a multi-agent setting, even in the presence of moving obstacles and complex dynamics.

\begin{figure} [t]
    \centering
    \includegraphics[width=0.85\linewidth]{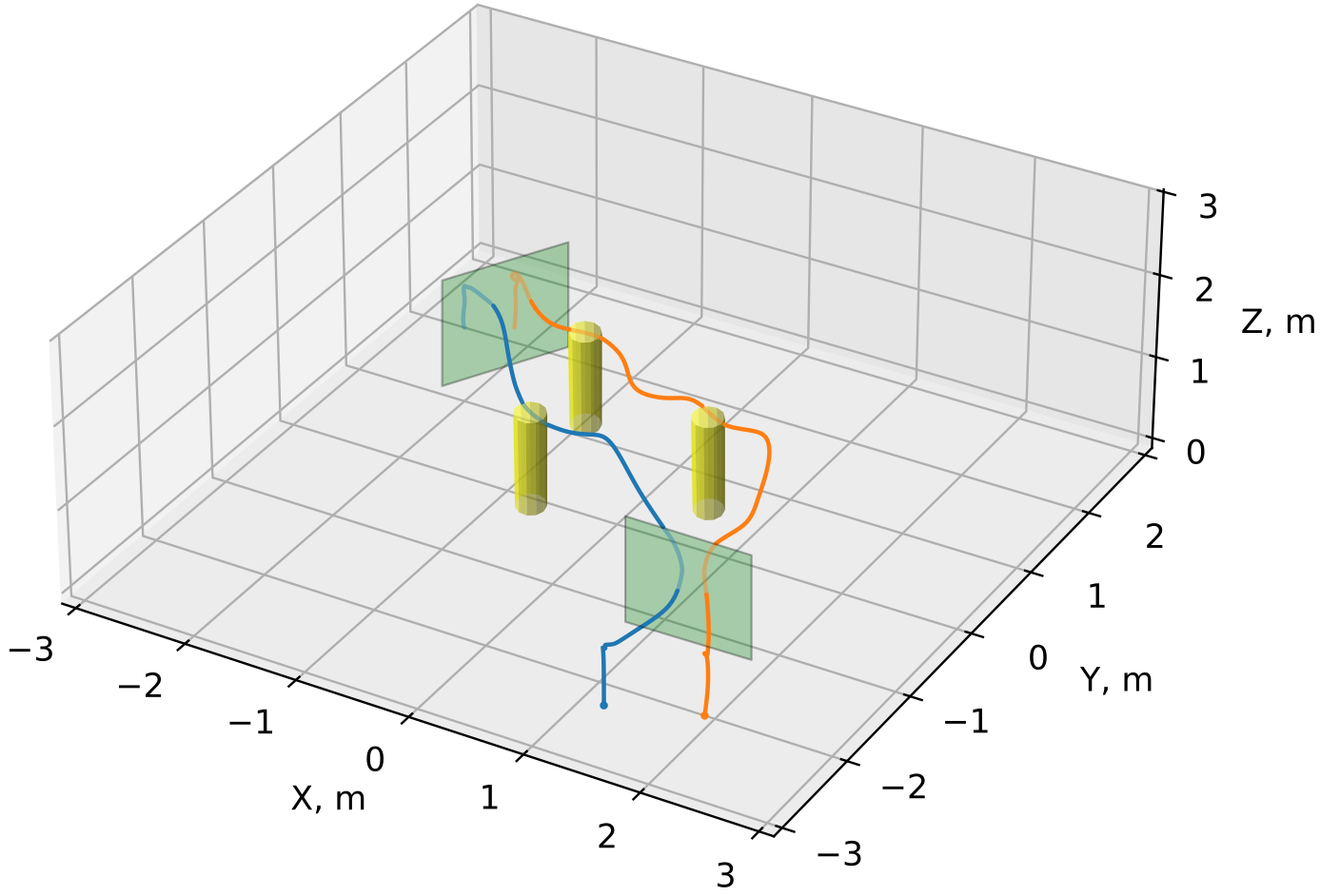}
    \caption{Recorded trajectories of two drones during a real-flight racing scenario. The drones navigate a course containing a green square gate and a yellow cylindrical obstacle, demonstrating agile flight and obstacle avoidance in a physical environment.}
    \label{fig:trajectory}
\end{figure}

We developed a centralized implementation of the MAPPO algorithm, combined with an Attention-based CBF NN. This attention mechanism ensures that the actions selected by the policy adhere to predefined safety constraints, such as maintaining a minimum safe distance from obstacles. Moreover, the NN structure is particularly effective for handling moving obstacles, as it dynamically adjusts the safe action set based on the current environment. 

\section{Experiments}
\subsection{Quantitative Evaluation of Swarm Navigation}

\begin{figure} [t]
    \centering
    \includegraphics[width=0.85\linewidth]{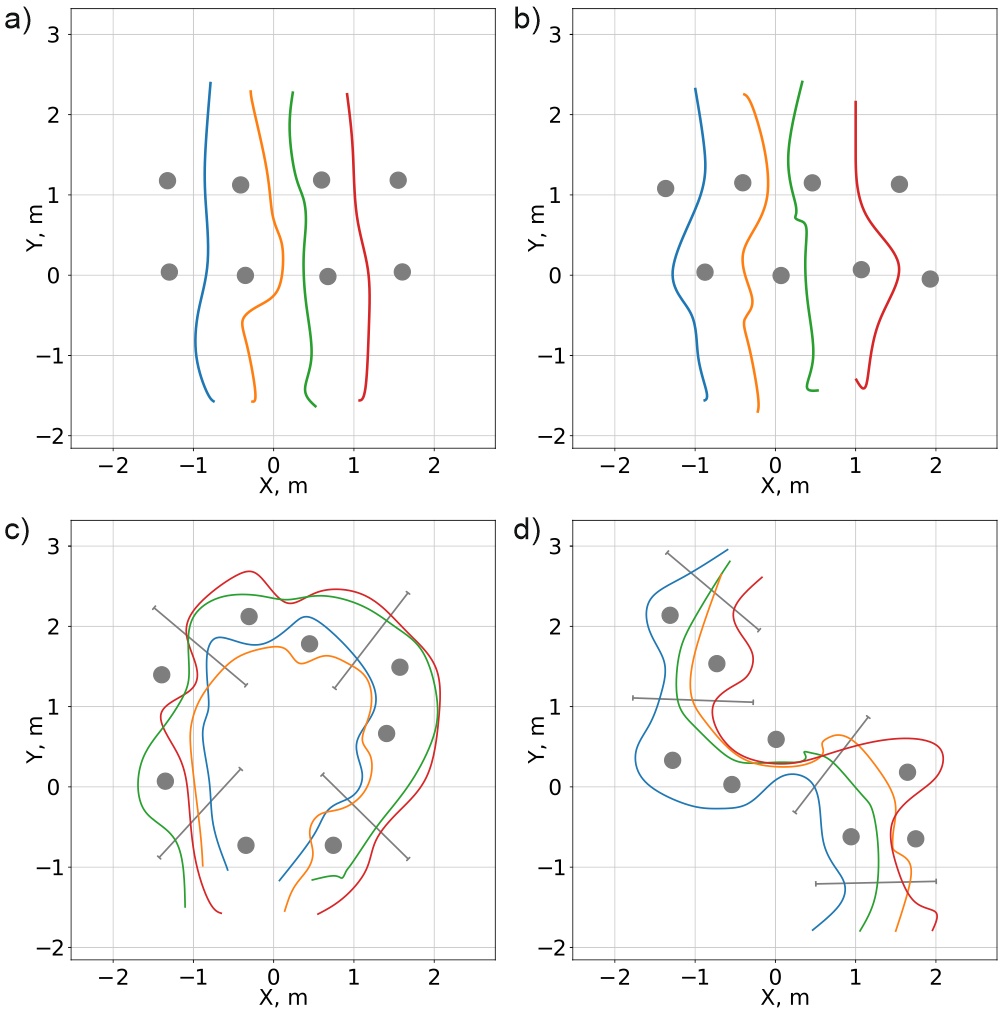}
    \caption{Evaluating swarm agility and path planning in complex simulation environments. Colored trajectories illustrate drones navigating two challenges: (a) and (b) flying straight through cluttered obstacles, and (c) and (d) coordinating in a complex racing track.}
    \label{fig:sim_exp_4_plots}
\end{figure}

To evaluate the performance of the proposed AttentionSwarm framework, we conducted 20 simulation trials under two distinct scenarios:

\begin{itemize}
  \item \textbf{Case~1}: Avoid the obstacles by a swarm of drones (Fig.~\ref{fig:sim_exp_4_plots}(a) and (b)).
  \item \textbf{Case~2}: Complete a race track with square gates while avoiding obstacles (Fig.~\ref{fig:sim_exp_4_plots}(c) and (d)).
\end{itemize}

Each trial involved a swarm of four drones operating in a shared 3D environment containing eight static obstacles.

In the case 1, the swarm achieved a 100\% mission success rate, maintaining a minimum clearance of 0.3~m from obstacles and an average inter-drone separation of 0.4~m. In the case 2, which included four gates and eight obstacles arranged along a closed circuit, the swarm achieved a 95\% completion rate. The average mission time was 20~s, corresponding to an average drone velocity of approximately 1.5~m/s. The final real-world experiment was conducted using two Crazyflie drones tasked with completing a racing track containing two gates and three obstacles. The swarm successfully accomplished the mission, demonstrating stable coordination and obstacle avoidance. The observed performance closely matched the simulation results, maintaining comparable accuracy and trajectory efficiency, as illustrated in Fig.~\ref{fig:trajectory}.

Table~\ref{table:attention_swarm_combined} presents a comparative evaluation of AttentionSwarm against standard baseline methods in a dynamic simulation environment. The proposed approach outperforms all baselines, achieving a perfect 100\% success rate with a single obstacle and maintaining 95\% with three obstacles. In contrast, conventional reinforcement learning methods such as MAPPO and A2C demonstrate significantly lower success rates, particularly as the environment complexity increases. The MPC baseline performs moderately but remains below AttentionSwarm. These results highlight the superior adaptability and robustness of the proposed framework in complex multi-obstacle scenarios.

\begin{table}[t]
\centering
\caption{Comparison of AttentionSwarm and Baseline Methods}
\label{table:attention_swarm_combined}
\begin{tabular}{|c| c| c| c|}
\hline

\textbf{Method} & 
\textbf{\begin{tabular}[c]{@{}c@{}}Case 1 \\ Precision (\%)\end{tabular}} &
\textbf{\begin{tabular}[c]{@{}c@{}}Case 2 \\ Precision (\%)\end{tabular}} &
\textbf{\begin{tabular}[c]{@{}c@{}}Collision\\ Rate (\%)\end{tabular}} \\

\hline
AttentionSwarm & 100 & 95 & 0 \\
MAPPO           & 50  & 40 & 30 \\
A2C             & 40  & 30 & 30 \\
MPC             & 60  & 50 & 15 \\
\hline

\end{tabular}
\end{table}

\subsection{Model Performance}
\begin{figure}[t]
    \centering
    \includegraphics[width=0.85\linewidth]{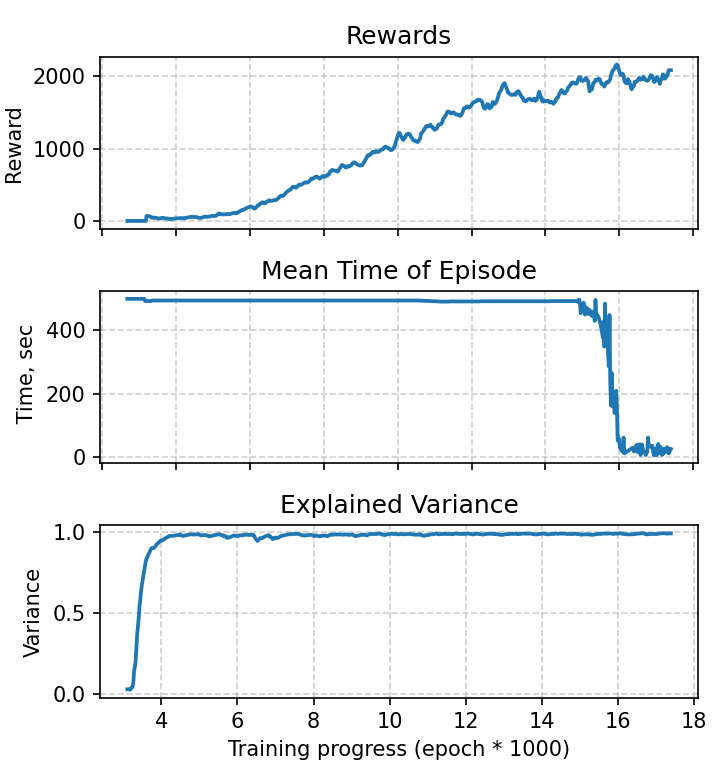}
    \caption{Convergence of the learning algorithm. The policy demonstrates stable learning, with rewards increasing and episode times shortening after 15,000 epochs, while the explained variance remains stable near 1.0, indicating successful training.}
    \label{fig:rewards}
\end{figure}

The policy was trained using the hyperparameters detailed in Tables \ref{tab:Hyperparameters} and \ref{tab:attention_params}. These settings enabled the stable convergence and performance improvements demonstrated in the learning curves. Fig.~\ref{fig:rewards} illustrates the training dynamics of the proposed neural network policy. The reward curve shows a consistent upward trend, indicating stable learning and progressive improvement in task performance. After approximately 3,000 training epochs, the policy begins to converge, achieving rewards exceeding 2000, which reflects successful optimization of the control policy. The mean episode time remains nearly constant during early training, followed by a sharp decrease after epoch 15,000, signifying that the agent learned to complete the mission more efficiently with reduced trajectory duration. The explained variance quickly rises toward 1.0 within the initial training phase and remains stable, confirming that the value function accurately models the underlying return distribution. 

\begin{table}[t]
\centering
\caption{Hyperparameters for MAPPO Training}
\label{tab:Hyperparameters}
\begin{tabular}{|@{}l|c|c@{}|}
\hline
\textbf{Parameter} & \textbf{Final Choice} \\
\hline
Actor LR & $5\!\times\!10^{-4}$ \\
\hline
Critic LR & $5\!\times\!10^{-3}$ \\
\hline
CBF LR   & $1\!\times\!10^{-3}$ \\
\hline
Clip Ratio $\varepsilon$ & 0.1 \\
\hline
GAE $\lambda$     & 0.95 \\
\hline
Discount $\gamma$  & 0.95 \\
\hline
Coefficient $\alpha$ for CBF loss function  & 0.1 \\
\hline
\end{tabular}
\end{table}

\begin{table}[t]
\centering
\caption{Attention Neural Network Parameters}
\label{tab:attention_params}
\begin{tabular}{|l|l|}
\hline
\textbf{Parameter} & \textbf{Value} \\ \hline
Input Embedding Dimension ($n_{\text{embed}}$) & 3 \\ \hline
Attention Module MLP Layers & 2 \\ \hline
Hidden Layer Dimensions & [1024, 256] \\ \hline
Activation Function & GeLU \\ \hline
\end{tabular}
\end{table}

\section{Conclusion and Future Work}

This study successfully developed and validated AttentionSwarm, a novel control framework for autonomous drone swarms in complex, obstacle-dense environments. Through extensive simulation and real-world experimentation, the proposed method demonstrated a superior combination of safety, efficiency, and robustness. In simulated environments, AttentionSwarm achieved near-perfect mission success rates (95-100\%) while reliably maintaining critical safety constraints for inter-drone separation and obstacle clearance. Its performance significantly surpassed established baseline methods like MAPPO, A2C, and MPC, particularly as environmental complexity increased. The efficacy of the simulated results was confirmed through real-world flights, where the swarm successfully navigated a physical race track with stable coordination and trajectory efficiency matching the simulations. Furthermore, the training dynamics confirmed the policy's stable convergence and its ability to learn increasingly time-optimal paths. Collectively, these results robustly establish AttentionSwarm as a highly promising solution for the deployment of autonomous swarms in real-world applications such as search and rescue, infrastructure inspection, and warehouse logistics, where reliable navigation in cluttered spaces is paramount.

\section*{Acknowledgements} 
Research reported in this publication was financially supported by the RSF grant No. 24-41-02039.






\balance

\bibliographystyle{IEEEtran}
\bibliography{references}

\begin{thebibliography}{10}
\providecommand{\url}[1]{#1}
\csname url@samestyle\endcsname
\providecommand{\newblock}{\relax}
\providecommand{\bibinfo}[2]{#2}
\providecommand{\BIBentrySTDinterwordspacing}{\spaceskip=0pt\relax}
\providecommand{\BIBentryALTinterwordstretchfactor}{4}
\providecommand{\BIBentryALTinterwordspacing}{\spaceskip=\fontdimen2\font plus
\BIBentryALTinterwordstretchfactor\fontdimen3\font minus \fontdimen4\font\relax}
\providecommand{\BIBforeignlanguage}[2]{{%
\expandafter\ifx\csname l@#1\endcsname\relax
\typeout{** WARNING: IEEEtran.bst: No hyphenation pattern has been}%
\typeout{** loaded for the language `#1'. Using the pattern for}%
\typeout{** the default language instead.}%
\else
\language=\csname l@#1\endcsname
\fi
#2}}
\providecommand{\BIBdecl}{\relax}
\BIBdecl

\bibitem{Tang_Abbatematteo_Hu_Chandra_Martín-Martín_Stone_2025}
\BIBentryALTinterwordspacing
C.~Tang, B.~Abbatematteo, J.~Hu, R.~Chandra, R.~Martín-Martín, and P.~Stone, ``Deep reinforcement learning for robotics: A survey of real-world successes,'' \emph{Proceedings of the AAAI Conference on Artificial Intelligence}, vol.~39, no.~27, pp. 28\,694--28\,698, Apr. 2025. [Online]. Available: \url{https://ojs.aaai.org/index.php/AAAI/article/view/35095}
\BIBentrySTDinterwordspacing

\bibitem{peter2024}
R.~Peter, L.~Ratnabala, D.~Aschu, A.~Fedoseev, and D.~Tsetserukou, ``Lander.ai: Drl-based autonomous drone landing on moving 3d surface in the presence of aerodynamic disturbances,'' in \emph{Proc. 2024 International Conference on Unmanned Aircraft Systems (ICUAS)}, 2024, pp. 295--300.

\bibitem{huang2024}
Z.~Huang, Z.~Yang, R.~Krupani, B.~Şenbaşlar, S.~Batra, and G.~S. Sukhatme, ``Collision avoidance and navigation for a quadrotor swarm using end-to-end deep reinforcement learning,'' in \emph{Proc. 2024 IEEE International Conference on Robotics and Automation (ICRA)}, 2024, pp. 300--306.

\bibitem{petertornado}
\BIBentryALTinterwordspacing
R.~Peter, L.~Ratnabala, D.~Aschu, A.~Fedoseev, and D.~Tsetserukou, ``Tornadodrone: Bio-inspired drl-based drone landing on 6d platform with wind force disturbances,'' \emph{ArXiv}, 2024. [Online]. Available: \url{https://arxiv.org/abs/2406.16164}
\BIBentrySTDinterwordspacing

\bibitem{wan2020}
W.~Wan, H.~Kim, N.~Hovakimyan, L.~Sha, and P.~G. Voulgaris, ``A safety constrained control framework for uavs in gps denied environment,'' in \emph{Proc. 2020 59th IEEE Conference on Decision and Control (CDC)}, 2020, pp. 214--219.

\bibitem{tao2022}
C.~Tao, H.~Kim, H.~Yoon, N.~Hovakimyan, and P.~Voulgaris, ``Control barrier function augmentation in sampling-based control algorithm for sample efficiency,'' in \emph{Proc. 2022 American Control Conference (ACC)}, 2022, pp. 3488--3493.

\bibitem{feng2025}
\BIBentryALTinterwordspacing
M.~Feng, V.~Parimi, and B.~Williams, ``Safe multi-agent navigation guided by goal-conditioned safe reinforcement learning,'' 2025. [Online]. Available: \url{https://arxiv.org/abs/2502.17813}
\BIBentrySTDinterwordspacing

\bibitem{xu2025}
\BIBentryALTinterwordspacing
Z.~Xu, X.~Han, H.~Shen, H.~Jin, and K.~Shimada, ``Navrl: Learning safe flight in dynamic environments,'' 2025. [Online]. Available: \url{https://arxiv.org/abs/2409.15634}
\BIBentrySTDinterwordspacing

\bibitem{safari2024}
\BIBentryALTinterwordspacing
A.~Safari and J.~B. Hoagg, ``Safe navigation in unmapped environments for robotic systems with input constraints,'' 2024. [Online]. Available: \url{https://arxiv.org/abs/2410.02106}
\BIBentrySTDinterwordspacing

\bibitem{zhao2023}
P.~Zhao, R.~Ghabcheloo, Y.~Cheng, H.~Abdi, and N.~Hovakimyan, ``Convex synthesis of control barrier functions under input constraints,'' \emph{IEEE Control Systems Letters}, vol.~7, pp. 3102--3107, 2023.

\bibitem{lapandic2021}
\BIBentryALTinterwordspacing
D.~Lapandić, L.~Persson, D.~V. Dimarogonas, and B.~Wahlberg, ``Aperiodic communication for mpc in autonomous cooperative landing,'' \emph{IFAC-PapersOnLine}, vol.~54, no.~6, pp. 113--118, 2021, 7th IFAC Conference on Nonlinear Model Predictive Control NMPC 2021. [Online]. Available: \url{https://www.sciencedirect.com/science/article/pii/S2405896321013070}
\BIBentrySTDinterwordspacing

\bibitem{andrew2022}
\BIBentryALTinterwordspacing
A.~Singletary, A.~Swann, Y.~Chen, and A.~D. Ames, ``Onboard safety guarantees for racing drones: High-speed geofencing with control barrier functions,'' 2022. [Online]. Available: \url{https://arxiv.org/abs/2201.04331}
\BIBentrySTDinterwordspacing

\bibitem{gu2023}
S.~Gu, J.~G. Kuba, Y.~Chen, Y.~Du, L.~Yang, A.~Knoll, and Y.~Yang, ``Safe multi-agent reinforcement learning for multi-robot control,'' \emph{Artificial Intelligence}, p. 103905, 2023.

\bibitem{bartolomei2022}
L.~Bartolomei, Y.~Kompis, L.~Teixeira, and M.~Chli, ``Autonomous emergency landing for multicopters using deep reinforcement learning,'' in \emph{Proc. 2022 IEEE/RSJ International Conference on Intelligent Robots and Systems (IROS)}, 2022, pp. 3392--3399.

\bibitem{wang2024}
\BIBentryALTinterwordspacing
Z.~Wang and N.~Mahmoudian, ``Vision-driven uav river following: Benchmarking with safe reinforcement learning,'' \emph{IFAC-PapersOnLine}, vol.~58, no.~20, pp. 421--427, 2024, 15th IFAC Conference on Control Applications in Marine Systems, Robotics and Vehicles CAMS 2024. [Online]. Available: \url{https://www.sciencedirect.com/science/article/pii/S2405896324018457}
\BIBentrySTDinterwordspacing

\bibitem{zhao2024}
G.~Zhao, T.~Wu, Y.~Chen, and F.~Gao, ``Learning speed adaptation for flight in clutter,'' \emph{IEEE Robotics and Automation Letters}, vol.~9, no.~8, pp. 7222--7229, 2024.

\bibitem{roohan2025}
R.~A. Khan, V.~Serpiva, D.~Aschalew, A.~Fedoseev, and D.~Tsetserukou, ``Agilepilot: Drl-based drone agent for real-time motion planning in dynamic environments by leveraging object detection,'' in \emph{Proc. 2025 International Conference on Unmanned Aircraft Systems (ICUAS)}, May 2025, pp. 185--192.

\bibitem{yifru2024}
L.~Yifru and A.~Baheri, ``Concurrent learning of control policy and unknown safety specifications in reinforcement learning,'' \emph{IEEE Open Journal of Control Systems}, vol.~3, pp. 266--281, 2024.

\bibitem{liu2024}
\BIBentryALTinterwordspacing
W.~Liu, H.~Zhao, C.~Li, J.~Biswas, B.~Okal, P.~Goyal, Y.~Chang, and S.~Pouya, ``X-mobility: End-to-end generalizable navigation via world modeling,'' 2024. [Online]. Available: \url{https://arxiv.org/abs/2410.17491}
\BIBentrySTDinterwordspacing

\bibitem{Amendola2024}
J.~Amendola, L.~R. Cenkeramaddi, and A.~Jha, ``Drone landing and reinforcement learning: State-of-art, challenges and opportunities,'' \emph{IEEE Open Journal of Intelligent Transportation Systems}, vol.~5, pp. 520--539, 2024.

\bibitem{srinivasan2020}
\BIBentryALTinterwordspacing
K.~Srinivasan, B.~Eysenbach, S.~Ha, J.~Tan, and C.~Finn, ``Learning to be safe: Deep rl with a safety critic,'' 2020. [Online]. Available: \url{https://arxiv.org/abs/2010.14603}
\BIBentrySTDinterwordspacing

\bibitem{yang2022}
T.-Y. Yang, T.~Zhang, L.~Luu, S.~Ha, J.~Tan, and W.~Yu, ``Safe reinforcement learning for legged locomotion,'' in \emph{Proc. 2022 IEEE/RSJ International Conference on Intelligent Robots and Systems (IROS)}, 2022, pp. 2454--2461.

\bibitem{Xie2023}
Z.~Xie and P.~Dames, ``Drl-vo: Learning to navigate through crowded dynamic scenes using velocity obstacles,'' \emph{IEEE Transactions on Robotics}, vol.~39, no.~4, pp. 2700--2719, 2023.

\bibitem{zhou2023}
Z.~Zhou, O.~S. Oguz, M.~Leibold, and M.~Buss, ``Learning a low-dimensional representation of a safe region for safe reinforcement learning on dynamical systems,'' \emph{IEEE Transactions on Neural Networks and Learning Systems}, vol.~34, no.~5, pp. 2513--2527, 2023.

\bibitem{tadevosyan2025}
G.~Tadevosyan, M.~Osipenko, D.~Aschu, A.~Fedoseev, V.~Serpiva, O.~Sautenkov, S.~Karaf, and D.~Tsetserukou, ``Safeswarm: Decentralized safe rl for the swarm of drones landing in dense crowds,'' in \emph{Proc. 2025 20th ACM/IEEE International Conference on Human-Robot Interaction (HRI)}, 2025, pp. 1665--1669.

\bibitem{Loquercio_2020}
A.~Loquercio, E.~Kaufmann, R.~Ranftl, A.~Dosovitskiy, V.~Koltun, and D.~Scaramuzza, ``Deep drone racing: From simulation to reality with domain randomization,'' \emph{IEEE Transactions on Robotics}, vol.~36, no.~1, pp. 1--14, 2020.

\bibitem{geles2024}
\BIBentryALTinterwordspacing
I.~Geles, L.~Bauersfeld, A.~Romero, J.~Xing, and D.~Scaramuzza, ``Demonstrating agile flight from pixels without state estimation,'' 2024. [Online]. Available: \url{https://arxiv.org/abs/2406.12505}
\BIBentrySTDinterwordspacing

\bibitem{zhang2025}
S.~Zhang, O.~So, K.~Garg, and C.~Fan, ``Gcbf+: A neural graph control barrier function framework for distributed safe multiagent control,'' \emph{IEEE Transactions on Robotics}, vol.~41, pp. 1533--1552, 2025.

\bibitem{zhang2024}
\BIBentryALTinterwordspacing
Z.~Zhang, H.~M.~S. Ahmad, E.~Sabouni, Y.~Sun, F.~Huang, W.~Li, and F.~Miao, ``Safety guaranteed robust multi-agent reinforcement learning with hierarchical control for connected and automated vehicles,'' 2024. [Online]. Available: \url{https://arxiv.org/abs/2309.11057}
\BIBentrySTDinterwordspacing

\bibitem{fisac2019}
J.~F. Fisac, A.~K. Akametalu, M.~N. Zeilinger, S.~Kaynama, J.~Gillula, and C.~J. Tomlin, ``A general safety framework for learning-based control in uncertain robotic systems,'' \emph{IEEE Transactions on Automatic Control}, vol.~64, no.~7, pp. 2737--2752, 2019.

\bibitem{mao2023}
\BIBentryALTinterwordspacing
Y.~Mao, Y.~Gu, N.~Hovakimyan, L.~Sha, and P.~Voulgaris, ``Sl1-simplex: Safe velocity regulation of self-driving vehicles in dynamic and unforeseen environments,'' \emph{ACM Trans. Cyber-Phys. Syst.}, vol.~7, no.~1, Feb. 2023. [Online]. Available: \url{https://doi.org/10.1145/3564273}
\BIBentrySTDinterwordspacing

\bibitem{jacquet2024}
\BIBentryALTinterwordspacing
M.~Jacquet and K.~Alexis, ``N-mpc for deep neural network-based collision avoidance exploiting depth images,'' 2024. [Online]. Available: \url{https://arxiv.org/abs/2402.13038}
\BIBentrySTDinterwordspacing

\bibitem{xiao2023}
W.~Xiao, T.-H. Wang, R.~Hasani, M.~Chahine, A.~Amini, X.~Li, and D.~Rus, ``Barriernet: Differentiable control barrier functions for learning of safe robot control,'' \emph{IEEE Transactions on Robotics}, vol.~39, no.~3, pp. 2289--2307, 2023.

\bibitem{11097824}
Y.~Yaqoot, M.~A. Mustafa, O.~Sautenkov, A.~Lykov, V.~Serpiva, and D.~Tsetserukou, ``Uav-vlrr: Vision-language informed nmpc for rapid response in uav search and rescue,'' in \emph{2025 IEEE Intelligent Vehicles Symposium (IV)}, 2025, pp. 1195--1200.

\bibitem{panerati2021}
J.~Panerati, H.~Zheng, S.~Zhou, J.~Xu, A.~Prorok, and A.~P. Schoellig, ``Learning to fly---a gym environment with pybullet physics for reinforcement learning of multi-agent quadcopter control,'' in \emph{Proc. 2021 IEEE/RSJ International Conference on Intelligent Robots and Systems (IROS)}, 2021, pp. 7512--7519.

\bibitem{ding2023}
\BIBentryALTinterwordspacing
D.~Ding, X.~Wei, Z.~Yang, Z.~Wang, and M.~R. Jovanović, ``Provably efficient generalized lagrangian policy optimization for safe multi-agent reinforcement learning,'' 2023. [Online]. Available: \url{https://arxiv.org/abs/2306.00212}
\BIBentrySTDinterwordspacing

\end{thebibliography}

\end{document}